\documentclass[manuscript]{aastex}

\newcommand{\kepler}{{\it Kepler}}

\newcommand{\prot}{\ensuremath{P_{\rm rot}}}
\usepackage{graphicx}
\usepackage{amsmath}
\usepackage{color}
\usepackage{natbib}
\usepackage{comment}
\usepackage[update,prepend]{epstopdf}
\usepackage{hyperref}

\shorttitle{Title}
\shortauthors{Mazeh, Peretz \& McQuillan, \& Goldstein}

\begin{document}

\title{Photometric Amplitude Distribution of Stellar Rotation of
{\it Kepler} KOIs---Indication for Spin--Orbit Alignment of Cool Stars
and High Obliquity for Hot Stars}

\author{
Tsevi Mazeh\altaffilmark{1,2},
Hagai B. Perets\altaffilmark{3},
Amy McQuillan\altaffilmark{1}
and
Eyal S. Goldstein\altaffilmark{1}
}

\altaffiltext{1}{School of Physics and Astronomy, Raymond and
Beverly Sackler Faculty of Exact Sciences, Tel Aviv University,
Tel Aviv 69978, Israel; \href{mazeh@post.tau.ac.il}{mazeh@post.tau.ac.il}}

\altaffiltext{2}{The Jesus Serra Foundation Guest Program, 
Instituto de Astrofísica de Canarias, C. via Lactea S/N, 38205 La Laguna, Tenerife, Spain}

\altaffiltext{3}{Physics Department, Technion: Israel Institute of Technology, Haifa 32000, Israel}

%-------------------------------

\begin{abstract}
The observed amplitude of the rotational photometric modulation of a star with spots should depend on the
inclination of its rotational axis relative to our line of sight.
Therefore, the distribution of observed rotational amplitudes
of a large sample of stars depends on the distribution
of their projected axes of rotation. 
Thus, comparison of the stellar
rotational amplitudes of the \kepler\ KOIs with those of \kepler\ single
stars can provide a measure to indirectly infer the properties
of the spin--orbit obliquity of \kepler\ planets. 
\\
We apply this technique to the large samples of 993 KOIs and 33,614 single \kepler\ stars in temperature range of 3500--6500\,K. We find with high significance that the amplitudes
of cool KOIs are larger, on the order of 10\%, than those of the
single stars.
In contrast, the amplitudes of hot KOIs are systematically lower. After correcting for an observational bias, we estimate that the amplitudes of the hot KOIs are {\it smaller} than the single stars by about the same factor of 10\%.
The border line between the relatively larger and smaller amplitudes, relative to the amplitudes of the single stars, occurs at about 6000\,K. \\
Our results suggest that the cool stars have their planets aligned with their stellar rotation, while the planets around hot stars have large obliquities, consistent with the findings of \citet{winn10} and \citet{albrecht12}.  We show that the low obliquity of the planets around cool stars extends up to at least 50 days, a feature that is not expected in the framework of a model that assumes the low obliquity is due to planet-star tidal realignment. 
\end{abstract}


%===================
\section{Introduction}
\label{sec:intro}
%===================

The photometry of the \kepler\ space mission \citep{borucki10}
are revolutionizing the study of
stellar rotation by providing four years of almost uninterrupted
light curves (LCs), with an unprecedented level of precision and
time sampling, for a very large sample of stars. The data
allow detection of photometric modulation induced by stellar spots that cross the stellar visible disk with the rotational period of the star. With the \kepler\ data we can detect stellar  
rotational periods even for slowly rotating
stars with low-amplitude modulation, for a well-defined
large sample
\citep[e.g.,][]{basri11, nielsen13, reinhold13}.
To derive the rotational periods from these data,
\citet{mcq+13a,mcq+14} used the auto-correlation function (ACF)
of each LC, which measures the degree of self-similarity
of the photometry over a range of time lags. 
This method yielded 34,030 rotational
periods for single main-sequence stars in the \kepler\ sample
\citep[][MMA14]{mcq+14}. By definition, 
the single-star sample of MMA14 did not
include any KOIs which showed evidence for transiting planets
\citep{batalha13}, nor detected eclipsing binaries
\citep[EB;][]{prsa11, slawson11}.

Given the large sample of MMA14, we can now compare the
observational statistical features of the stellar rotation of the single
stars of \kepler\ with the ones of the KOIs.
\citet[][MMA13]{mma13b} compared the stellar rotational {\it
periods} of the KOIs with those of the single stars, while this
work considers the observed photometric {\it amplitudes} of the
two groups.

The observed rotational amplitude of a star
(e.g., \citealt{basri11, mcq+13a, walkowicz13}; MMA14) 
depends on the inclination of its
rotational axis relative to our line of sight---stars with low rotational inclinations appear with lower observed amplitudes \citep[e.g.,][]{jackson12}. 
Therefore, the distribution of the observed rotational amplitudes
of a sample is not the real-amplitude distribution, as measured
by an observer with a line of sight orthogonal to the stellar
rotational axes. Instead, subject to an observational detection threshold, we measure the real distribution convolved with some spread function, which
reflects the inclination distribution of the sample and the
dependence of the amplitude on the inclination \citep[e.g.,][]{jackson13}.

These considerations suggest that we can have some indirect
access to the distribution of stellar inclinations of the
\kepler\ KOIs relative to the inclinations of the single stars by
comparing their observed amplitude distributions, if we assume
that the distributions of their real amplitudes are the same.
Naively, we could expect the stellar inclinations of the KOIs to
be close to $90^{\circ}$, and therefore the
rotational amplitudes of the KOIs to be larger than those of
the \kepler\ single stars, provided 
{\it the KOI's rotational axes are aligned with the orbital inclinations of their transiting planets}, which are
almost orthogonal to our line of sight.  

The angle between the stellar spin axis and the orbital planetary angular momentum for an individual planet, also
referred to as  the obliquity of the system, is
a matter of intense study in recent years.  
It typically relies on 
spectroscopic measurements during transit, making use of the Rossiter--Mclaughlin (R-M) effect \citep{queloz00} to measure the projected obliquity on the sky
\cite[e.g.,][]{triaud10, moutou11, winn11, albrecht12}. This was done 
primarily for hot Jupiters---gas-giant planets at short-period orbits. The extensive effort has 
lead to measurements of a few tens of spin--orbit projected inclinations.

The line-of-sight component of the spin--orbit angle can be
measured using
asteroseismology \citep{gizon03, chaplin13}, or the observed
stellar rotational line broadening, {\it if} the
host star radius and rotational period are known with sufficient
precision \citep{hirano12, hirano14, walkowicz13}. 
\citet{schlaufman10} has taken a statistical approach to identify KOIs with small stellar line broadening with respect to its expected distribution, if the stars were aligned with the planetary motion, given the stellar type and the experimental errors. He found 10 systems with probable high obliquity. 
Finally, at about the submission of this paper, \citet{morton14} published an analysis of the observed rotational line broadening and rotational periods of 70 KOIs, obtaining posterior probability of the distribution of the stellar obliquities. They suggested  with a 95\% confidence level 
that the obliquities of stars with only a single detected transiting planet are systematically larger than those with detected multiple transiting planets.

An interesting method was applied by \cite{nutzman11} and \cite{sanchis11a}, who
used brief photometric signals during transit induced by the
transiting object moving across spots located on the surface of
the host object.
 This method required identification of the ``spot-crossing" event
within a transit, and has been applied to additional systems using short-cadence \kepler\ LCs
\citep[e.g.,][]{sanchis13}.

In a seminal work \citet{fabrycky09} noticed that some of the transiting planets were found to be aligned in a prograde orbit, with obliquity close to zero, while others were found to be
misaligned, including systems in retrograde motion, where the
spin--orbit angle is close to $180^\circ$ \citep[e.g.,][]{winn10, hebrard11}.
\citet{fabrycky09} inferred the existence of two populations of hot
Jupiters---well-aligned with the stellar rotational axes and isotropic.  
\citet{winn10} discovered
that hot stars host high obliquity systems, while cool stars have their hot Jupiters all aligned. As shown also  by \citet{albrecht12}, the transition between the two populations occurs at effective temperature,  $T_{\rm eff}$, of 6250\,K. 
A second obliquity trend was suggested by \citet{hebrard11}, who pointed out that 
 retrograde planetary motion was observed only for low-mass planets, below 3 Jupiter masses \citep[see a concise summary of the observational evidence for the obliquity distribution by][]{dawson14}. 

All these observational trends are crucial for our understanding the formation and dynamical evolution of planetary systems.  We 
therefore set out to use the derived {\it amplitudes} of the stellar rotational photometric modulations of the KOIs and compare them to the amplitudes of the single stars observed by \kepler . Our goal is to confirm or refute some of the statistical trends reported, the alignment of the cool star axes with their planets in particular, and to explore the range of orbital periods for which the alignment can be traced.
 
Section~\ref{sec:data} describes the target selection for the KOIs
and Section~\ref{sec:rot_meas} the derivation of the rotational periods and
their amplitudes. We then compare the KOI amplitudes to those of
the single-star sample of MMA14 in Section~\ref{sec:comp}, and perform extensive simulations to study some possible observational biases in Section~\ref{sec:selection}. Section~\ref{sec:geometry} shows that a natural interpretation of our findings is that planets around cool stars have their orbits aligned with the rotational axes of their parent stars, while the hot star systems have large obliquities. Section~\ref{sec:period} shows that the low obliquity of the cool stars extends up to at least 50 days. Section~\ref{sec:summary} summarizes the results of this study, and points out the possible implication of the findings on the models of planet obliquities.

%====================
%Section 2
\section{KOI Sample}
\label{sec:data}
%=====================

For this work we used the list of the KOIs and their parameters
from the NASA Exoplanet
Archive\footnote{http://exoplanetarchive.ipac.caltech.edu}
(\citealt{akeson13}; NEA)
%\citep[][NEA]{akeson13}
of 2014  March 25. Targets identified as false
positives by either the \kepler\ pipeline or the NEA were
excluded, leaving 3685 KOIs.

Targets listed as EBs on the Villanova Eclipsing Binary web
site\footnote{http://keplerebs.villanova.edu} were removed from
the sample. 
Admittedly, some of the systems in the EB catalog are probably giant, sometimes inflated, planets. Kepler-76 \citep{faigler13} is one example. Nevertheless, we preferred to lose a small number of planets than contaminating our sample with many EBs.  
The EB website hosts an up-to-date and extended version
of the \kepler\ EB catalogs of \cite{prsa11} and
\cite{slawson11}. The EB catalog used in this work was downloaded
on 2014 Apr 1, and included 302 targets in our KOI
sample.
Five stars had neither \kepler\ Input Catalog (KIC) or NEA $T_{\rm eff}$
available so  were also removed from the sample.
We were left with 3355 host stars,
listed in Table~\ref{tab:pers}
\footnote{http://wise-obs.tau.ac.il/\~{}mazeh/Table1\_StellarRotationsOfKOIs.txt}, with their periods and amplitudes when available.
%%%%%%%%%%%%%%%%

The effective temperatures and surface gravities used in this
work come from the KIC \citep{brown11}
or, when available, from \cite{dressing13}. Targets with parameters
from \cite{dressing13} have the ``D" flag set in
Table~\ref{tab:pers}. For $121$ stars in the
sample that are missing KIC values for $\log g$ and $T_{\rm
eff}$, we used those provided in the NEA. These objects have the
``N" flag set in Table~\ref{tab:pers}.

We performed additional binarity checks and identified eight KOIs
as likely EBs due to depth differences in alternate transits.
These are flagged as ``F" in Table~\ref{tab:pers} and excluded from the
study, leaving 3347 targets.

We used the $\log g$ and $T_{\rm eff}$ cut of \cite{ciardi11} to
identify 138 likely giants, which have the ``G"
flag set in Table~\ref{tab:pers}. Three of these were previously
excluded as
likely false positives from this work (flag ``F"). The likely
giants were excluded from the analysis
because we focus on main-sequence stars only.

The temperatures of most stars in our KOI sample were found in the
range of $3500$--$6500$\,K.
%Outside this range, the KOIs density per temperature is quite
%sparse ---
Only 124 stars hotter than or equal to 6500\,K or cooler
than or equal to 3500\,K are listed in Table~\ref{tab:pers}.  We have opted
to exclude  them from the analysis, in order to enable us to
concentrate on a well defined temperature range that included most of
the stars.
% In this study we wish to focus only on stars with definite
The rejected KOIs are noted by the ``T" flag
in Table~\ref{tab:pers}. Two of these were previously excluded as
likely false positives (flag ``F") and one as a likely giant (flag
``G"). This provides a total of 3,091 stars to
which the automated autocorrelation function (AutoACF)
was applied.

%=========================
%Section 3
\section{Rotational Period Measurement}
\label{sec:rot_meas}
%=========================

\citet{mma13b} derived already the rotational periods of 737 KOIs. Since then the list of KOIs and EBs were updated and more quarters became publicly available. We therefore performed the search for rotational photometric modulation again, using now the public release 14 quarter 3--14 (Q3--Q14) LCs, which were downloaded from the \kepler\ mission archive.\footnote{http://archive.stsci.edu/kepler} 
We omitted Q0
and Q1 due to their short duration and Q2 because of
significant residual systematics. We used the data corrected for
instrumental systematics using PDC-MAP
%\citep{smi+12, stu+12}
\citep{smith12,stumpe12}, which removes the majority of
instrumental glitches and systematic trends using a Bayesian
approach, while retaining most real (astrophysical) variability.
Of the 3685 KOIs, 3662 had publicly available PDC-MAP Q3-Q14 LCs.  

The rotational period measurement was performed using the AutoACF technique, described in detail
in MMA14. This method is based on the measure of the
degree of self-similarity of the LC over a range of
lags. In the case of rotational modulation, the repeated
spot-crossing signatures lead to ACF peaks at lags corresponding
to the rotational period and integer multiples of it. The basic
pre-processing steps applied to the PDC-MAP LCs are
described in \citet{mcq+13a, mma13b}, including median
normalization of each quarter and masking transit events.

The automation of the detection method uses a training set of
visually verified ACF results to select only period
detections with high probability of being true rotational
measurements, based on the ACF peak height, its period and the
stellar temperature
$T_{\rm eff}$, and the consistency of the detection in different
sections of the LC. A detailed description is given by
MMA14.
KOIs with inconsistent detections in different quarters were flagged by 
``M1" in the table, and the ones with peaks not high enough, depending on the temperature and period, by ``M2''. 
The AutoACF  yielded 1031 period detections.
Since there are a small number of KOIs compared to the full sample for which
the automation routine was developed, we also visually examined
all AutoACF periodic KOI detections, and removed 19
targets which either had false period detections, or the period
did not appear to arise from rotational modulation. These have
the ``R" flag set in Table~\ref{tab:pers}.

The visually verified AutoACF analysis yielded 1012 period detections.
We found that for 18 of these targets, the centroid
motions on the \kepler\ CCD indicate that the transits and rotation are on different targets (flag
``C"). A single star, KIC\,8043882, was removed from the sample
because
a valid amplitude of modulation measurement could not be
made due to missing data sections.

Once these potential contaminants were removed, we were left
with 993 period detections for main-sequence
planet hosts, upon which this study focused.
As in MMA14, to derive the photometric amplitude we divided the
LC of each KOI into time bins of the detected rotational
period, obtained the photometric difference between the
5th and the 95th percentiles of normalized flux for all time bins, and
then took the median of these values.
The detected periods were between $0.2$ and $65$\,days,
and the amplitude of the modulation, $R_{\rm var}$, ranges from
$0.1$ to $100$\,mmag.

%----------------------------------------------------------------
%
%  Table 1
%
\begin{table*}
  \caption{Stellar Rotational Periods for the KOIs}
  \label{tab:pers}
  \centering
  \begin{tabular}{ccccccccc|c}
  \multicolumn{9}{c|}{Stellar and Rotational Detection Parameters}
& \multicolumn{1}{|c}{Flags\tablenotemark{a}}\\
  \hline
    KOI   &   KIC   &   $T_{\rm eff}$   &   $\log g$   &
$\prot$   &
    $\sigma_{\rm  p}$    &    $R_{\rm
var}$   &    LPH  & $w$ &
    \dots \\
    &   & (K) & (g/cm$^3$)  & (days)  & (days) & (ppm) & &\\
      \hline\hline
      3 & 10748390 & 4766 & 4.59 & 29.319 & 0.497 & 12246 &
0.61 & 0.43 & \dots \\
      42 & 8866102 & 6170 & 4.10 & 20.851 & 0.091 & 1113 &
1.02 & 0.52 & \dots\\
      49 & 9527334 & 6000 & 4.50 & 8.594 & 0.030 & 15673 &
1.23 & 0.68 & \dots \\
      82 & 10187017 & 4908 & 4.70 & 26.407 & 0.485 & 4053 &
0.72 & 0.49 & \dots \\
      85 & 5866724 & 6006 & 4.07 & 7.882 & 0.210 & 269 & 0.68
& 0.46 & \dots \\
      \hline\\
  \end{tabular}
  \tablecomments{
   $^{\rm a}$ Flags included in the full online table have the following
definitions:
D: values of \cite{dressing13} used, N: KIC values
missing so NEA values used; C: centroid motion shows
transit and rotational modulation on different stars; G: likely
giant; T: $T_{\rm eff}$ outside the range 3500--6500\,K; 
F: FP (EB) identified in this work; R: rejected by visual
examination; M1: inconsistent detections in different quarters; M2: correlation peaks not high enough, depending on temperature and period.
\\
 (This table is available in its 
entirety in a machine-readable form at http://wise-obs.tau.ac.il/\~{}mazeh/Table1\_StellarRotationsOfKOIs.txt.)}
\end{table*}
%----------------------------------------------------------------

%============================
%Section 4
%
\section{Comparison with \kepler\ Single Stars}
\label{sec:comp}
%============================

To compare the KOIs with the \kepler\ single-star sample we use the results of MMA14
who reported rotational periods for 34,030 main-sequence stars
observed by \kepler.
The selection criteria for these targets were described in detail
in MMA14,
and included the same $T_{\rm eff}$ and $\log g$ cuts for
main-sequence as are used in the
present paper, as well as removal of known KOIs and EBs.
For the present work we constrain the $T_{\rm eff}$ range to
3500--6500\,K, leaving a total sample
of 33,614 single stars.

\begin{figure*}[!h]
  \centering
  \includegraphics[width=1.1\linewidth]{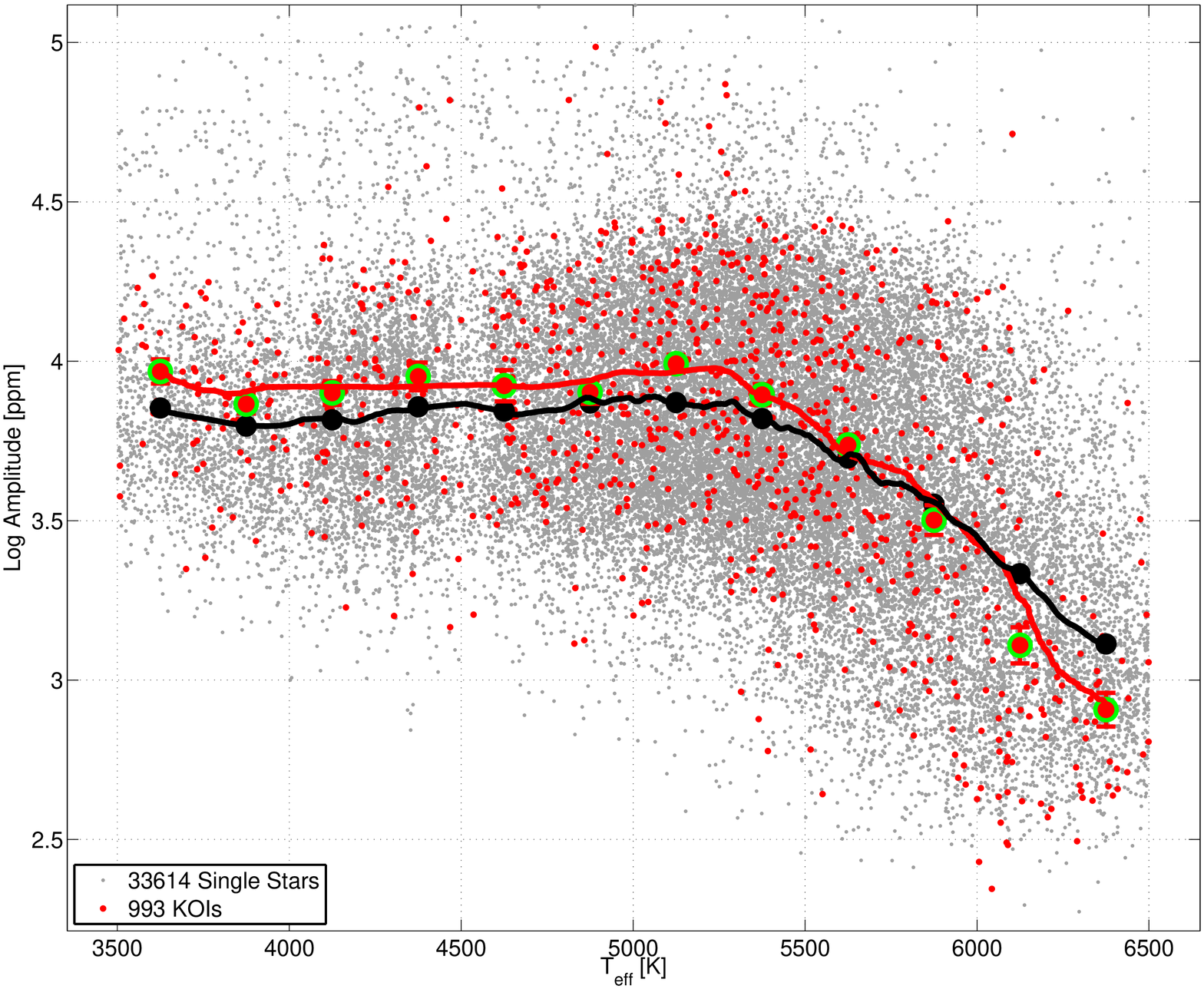}
  \caption{
Derived amplitudes of the photometric stellar rotation for
\kepler\ single stars and KOIs as a function of stellar temperature. 
There are 33,614 (black) single stars
and 993 (red) KOIs in the figure. Amplitudes are given by their
log values in ppm. Each of the two samples are divided into
250\,K temperature bins, and the median of each bin is plotted.
The error on the median is the ${\rm
MAD}/\sqrt{n}$ of the sample in each bin, where $n$ is the number of 
points per bin. For most bins, the error bars are smaller than the points. Also plotted are the running medians of each sample (see text).
}
\label{fig:AmpTemp}
\end{figure*}

To compare the amplitudes of photometric modulation of the KOIs to
those of the single stars of the \kepler\ sample we plot in
Figure~\ref{fig:AmpTemp} the amplitudes of the stars in the two samples as a function of the stellar temperature.  In addition, each of the two
samples are divided into 250\,K temperature bins, and the median
of each bin is plotted. The scatter of each bin is estimated by its MAD---the median of the absolute deviation of the amplitudes with respect to the  median of each bin, multiplied by 1.48. We prefer the MAD estimate as it is less sensitive to the outliers.
The error on the median of each bin, also plotted in the figure, 
is  the ${\rm MAD}/\sqrt{n}$ of the sample in 
each bin, where $n$ is the number of points per bin. The figure also displays for each sample a running median, with 1000 and 250 points width for the single and the KOIs amplitudes,  which were then smoothed with a width of 501 and 51 points, respectively. 

%----------------------------------------------------------------
% Figure 1
%
%----------------------------------------------------------------

\begin{figure*}
  \centering
  \includegraphics[width=\linewidth]{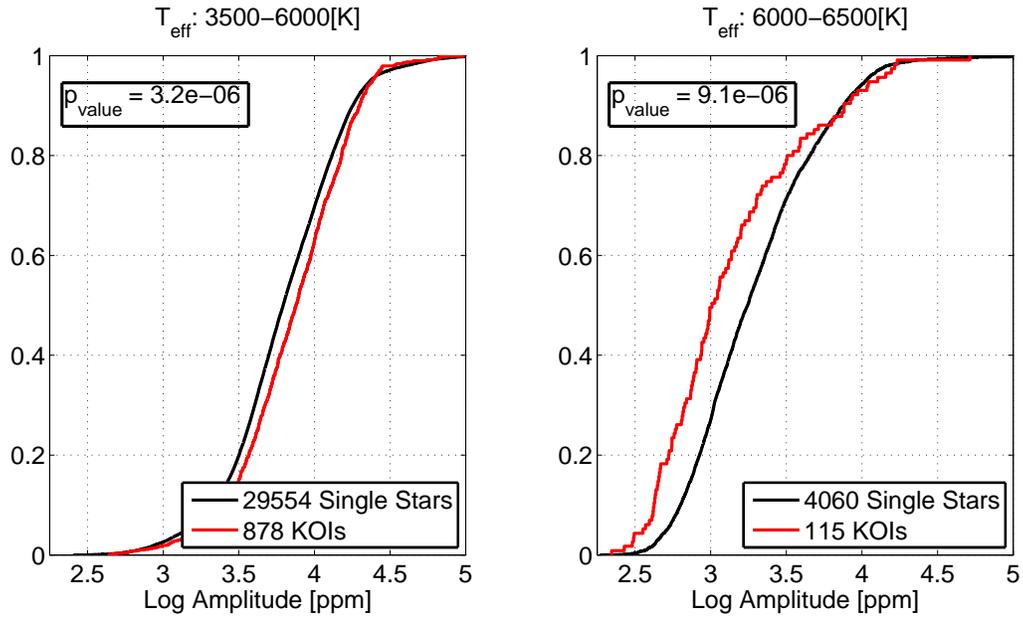}
  \caption{
Cumulative distributions of the photometric amplitudes of the \kepler\ single stars and the KOIs for two ranges of stellar temperatures. The
left panel shows the distributions of the amplitudes
of the cool stars, at range of 3500--6000\,K. Red is for the KOIs
and black for the single stars.  The right panel shows the
distributions for the hot stars, 6000--6500\,K. The {\it p}-value of the two-sample K-S test for each range of temperatures is also given.
}
\label{fig:AmpKS}
\end{figure*}

Figure~\ref{fig:AmpTemp} suggests that the KOIs have typically
larger amplitudes than the single stars for all but the hottest two
$T_{\rm
eff}$  bins, 6000--6500\,K, which show significantly lower
amplitudes for the KOIs. To check the significance of the
differences we
show in Figure~\ref{fig:AmpKS} the results of a
two-sample Kolmogorov--Smirnov (K-S) test for the KOIs and the
single stars, performed for separate two temperature ranges---3500--6000
and 6000--6500\,K.
The {\it p}-values for the null hypothesis, which assumes that the KOIs
and the single stars are both drown from the same underlying
amplitude distribution, are shown above the two
panels of Figure~\ref{fig:AmpKS}. The {\it p}-values of the two-sample K-S test, $3.2\times10^{-6}$ and $9.1\times10^{-6}$, for the two temperature bins, respectively, suggest that
the differences between the KOIs and the single stars are highly
significant in both ranges of temperatures.

%----------------------------------------------------------------
% Figure 2
%
%----------------------------------------------------------------

%============================
%Section 5
%
\section{Selection Effects?}
\label{sec:selection}
%============================

The amplitude distributions of the KOIs and the single stars seem significantly different. However, before we can conclude that this is indirect evidence for a small star--planet obliquity of the cool stars and a high obliquity of the hot stars we have to consider possible selection effects that could have caused the amplitude differences. We consider in this section two possible such observational biases, one that has to do with the rotational periods of the KOIs and the single stars, and the other with stellar noise of the hot stars that determines the transit detection threshold. 

To consider the period issue, we constructed in each temperature bin a series of single-star subsamples with rotational period distributions similar to that of the KOIs,  and showed that we obtain the same amplitude difference as with the real single-star sample. To study the problem of the stellar noise and the detection threshold, we generated a series of randomly simulated hot KOI samples that could have been detected, and compare their amplitudes with those of the real KOI and single-star samples. We find that the detection bias introduces a decrease in the KOI's amplitudes, but the actual sample of KOIs displays a difference too large to be accounted for by the selection effect and is probably due to high obliquity. 

\subsection{Rotational Periods}
%-----------------------------------

Rotational periods and amplitudes are known to be correlated
(e.g., \citealt{piz+03, har+11}; MMA14), with typically higher amplitude
variability at shorter periods. Therefore, the amplitude difference we observe could be a result of a difference between the rotational periods of the KOIs and those of the single stars. To face this problem we performed a follow-up test of comparison, with
matching period distributions for the KOI and the single-star
samples,
to remove any possible bias introduced by different period
distributions.
We divided the sample into bins of 500\,K,
selecting for each $T_{\rm eff}$-bin a {\it random} sample of single
stars that  approximately matched the
period distributions of the KOIs of that temperature, and compared the median amplitudes of the KOIs and the single stars in each bin.

\begin{figure*}[h!]
  \centering
  \includegraphics[width=1.1\linewidth]{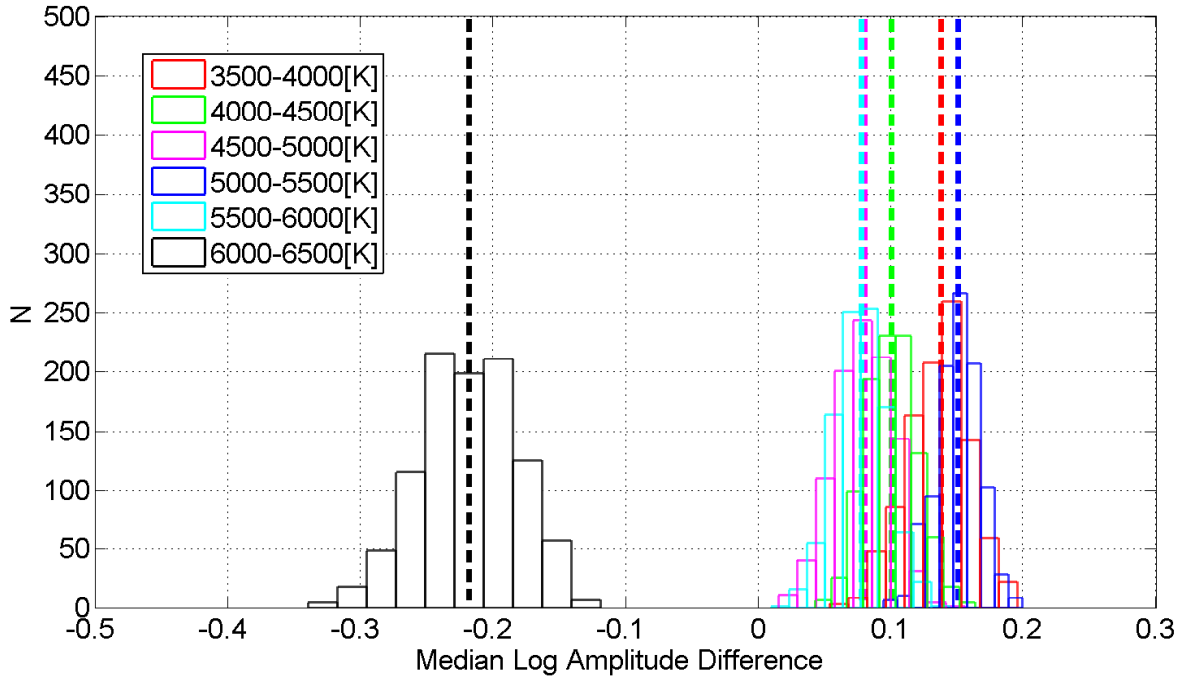}
  \caption{
Differences between the median log-amplitudes of the KOIs and the
single stars {\it with the same period distribution} for several temperature bins. 
Histograms show the differences of the log-amplitude medians derived for 1000 random  (see text) single-star subsamples.
The median of the differences for each $T_{\rm eff}$ bin is
shown by a vertical dashed line. The colors present the different temperature bins.
}
  \label{fig:AmpDiffHist}
\end{figure*}

This was done by
dividing the period distributions into $10$ bins of width $0.25$
in log period between $-0.5$ and $2$, in days, with two
additional bins for periods outside this range. We then selected
a random sample of single stars, with twice the number of stars
in each period bin as were in the KOIs sample for the same period
bin. This was the maximum integer multiple possible in all
$T_{\rm eff}$ and period bins. We then derived for each random
choice of the single-star sample its median value.
This process was then repeated 1000 times, with a new
random sample of non-KOIs drawn each time, and the same full
sample of KOIs used in each run.
For each random sample we derived the difference between the
median
of the log amplitudes of the KOIs and the single stars, obtaining
for each temperature bin a distribution of differences of medians,
which were plotted in Figure~\ref{fig:AmpDiffHist}.

%----------------------------------------------------------------
%Figure 3
%
%----------------------------------------------------------------

Figure~\ref{fig:AmpDiffHist} is consistent with the picture
revealed by the two previous ones. For the cool stars, the
KOIs have, on the average, larger amplitudes than the
single ones, on the order of $0.1$ in log amplitude, or a factor of
$\simeq 1.25$ larger. The hot KOIs, on the other hand, have
smaller amplitudes, on the order of $0.2$ in log
amplitude or a factor of $\simeq 0.6$ smaller. 

We therefore conclude that the amplitude difference we observe is {\it not} a result of a period difference between the KOIs and the single stars.

\subsection{Stellar Noises and the Hot Stars}

As pointed out by the anonymous referee, the small amplitudes of the hot stars could have been the result of observational selection effect. 
In his/her wise words, ``planets are more challenging to detect around stars with high variability, causing their stellar variability to appear systematically lower. This could be an issue for the hot stars which, on the main sequence, are larger than cool stars, making it more difficult to detect ... planets except around the least variable stars". 

In order to estimate the magnitude of the contribution of the  selection effect induced by the planet detection, following the referee's suggestion, we have generated 100,000 {\it simulated} samples of ``planets around hot stars", for which their parent stars where randomly chosen under \kepler\ discovery constraints. The goal is to find out how many of the simulated samples showed small amplitudes like the real sample, in order to estimate the significance of the amplitude difference between the KOIs and the hot single stars. 

To form a simulated sample  associated with the observed planets we have considered two real samples---the sample of planets orbiting hot stars with detected rotational periods (115 systems; Sample $\mathcal{P}$) and the sample of all hot stars with known rotational periods and amplitudes, including the KOIs with known periods (4147 systems; Sample $\mathcal{R}$). 
For each planet \{$P_i \in \mathcal{P}, i=1:115$\} we constructed a subsample, $\mathcal{R}_i$, chosen from the large sample $\mathcal{R}$, consisting of stars that could have been the parent star of $P_i$ and still be detected, given the stellar radius and photometric noise. 
A simulated sample of KOIs was then constructed by choosing randomly a star from each of the subsamples \{$\mathcal{R}_i, i=1:115$\}. 

For a star to be considered as a potential parent star to a planet $P_i$ and therefore included in the  $\mathcal{R}_i$ simulated sample, we required the potential  star to 
have  signal-to-noise ratio S/N$\geq10$, had it had the $P_i$ around it, 
in case the actual transit \{S/N\}$_i \geq 10$, and S/N$\geq \{{\rm S/N}\}_i$, if the actual  transit \{S/N\}$_i < 10$.

To derive the potential S/N we used
 \citet{howard12} expression, which we write here as: 

\begin{equation}
{\rm S/N}=\frac{R_{\rm pl}^2}{{\sigma}_{\rm CDPP}R_*^2}
 \sqrt{\frac{n_{\rm tr}\cdot t_{\rm dur}}{3\, {\rm hr}}} \ ,
\end{equation}
where $n_{\rm tr}$ is the number of observed transits, $t_{\rm dur}$ is the transit duration, $R_*$  and $R_{\rm pl}$ are the stellar and planet radii and $\sigma_{\rm CDPP}$ is the 3-hr photometric noise of \kepler. In fact, \kepler\ site includes such a noise estimate for each observed quarter, so we used the median of these values, $\bar{\sigma}_{\rm CDPP}$, for our simulations.  The stellar radius was estimated from the stellar mass, $M_*$, derived by \citet{mcq+14}, assuming main-sequence $R_*\propto M_*^{0.8}$ relation. We also assume that $t_{\rm dur}\propto R_*$, and therefore conclude that for a given planet

\begin{equation}
{\rm S/N} \propto\frac{1}{\bar{\sigma}_{\rm CDPP}R_*^{3/2}} 
 \ .
\end{equation}

\begin{figure*}[h!]
  \centering
  \includegraphics[width=1.1\linewidth]{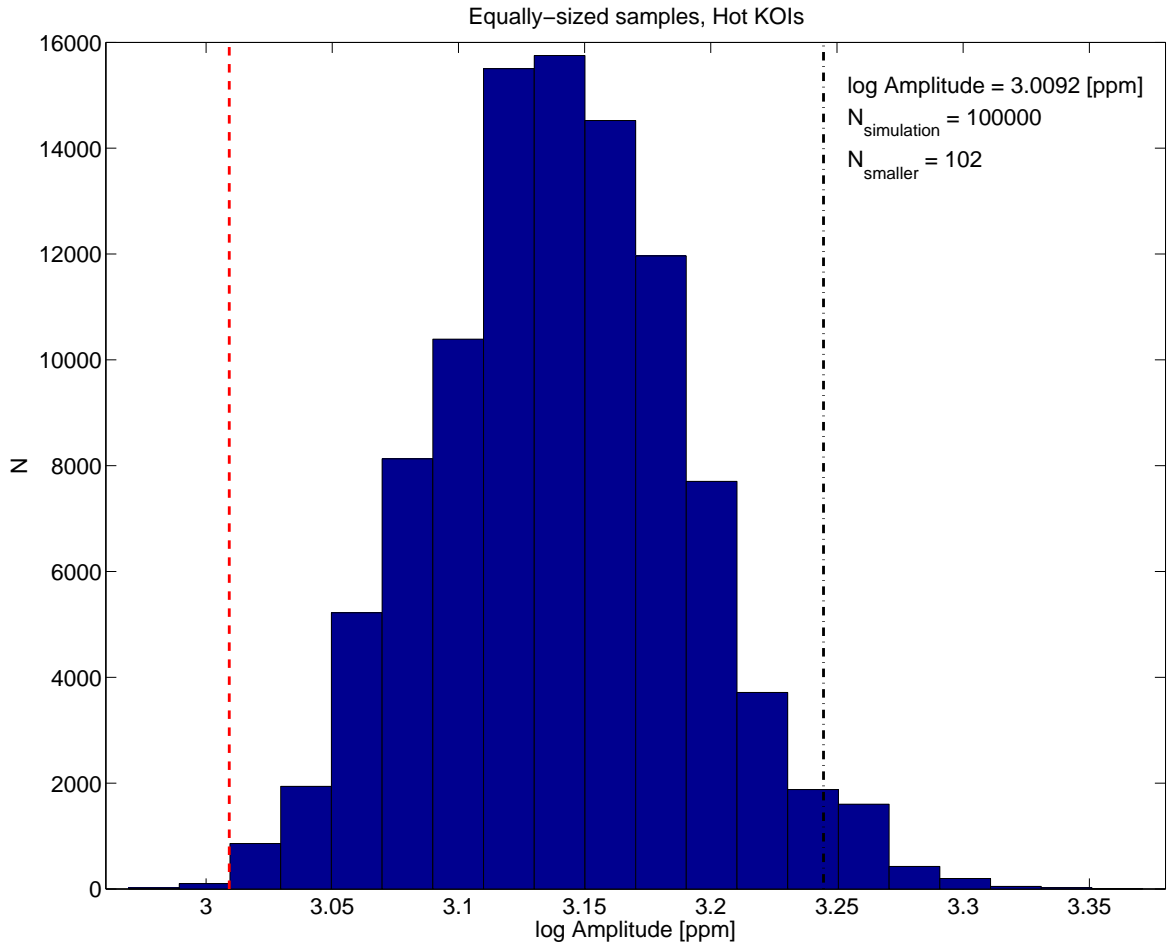}
  \caption{
Histogram of the median values of the amplitudes of 100,000 simulated samples of 115 planets around hot stars (see text). Also marked are the value of the real sample (in dashed red line) and median of all the hot stars (in dashed-dot blue line). 
}
\label{fig:SimulationAmpMedianHot}
\end{figure*}

\begin{figure*}[h!]
 \centering
 \includegraphics[width=1.1\linewidth]{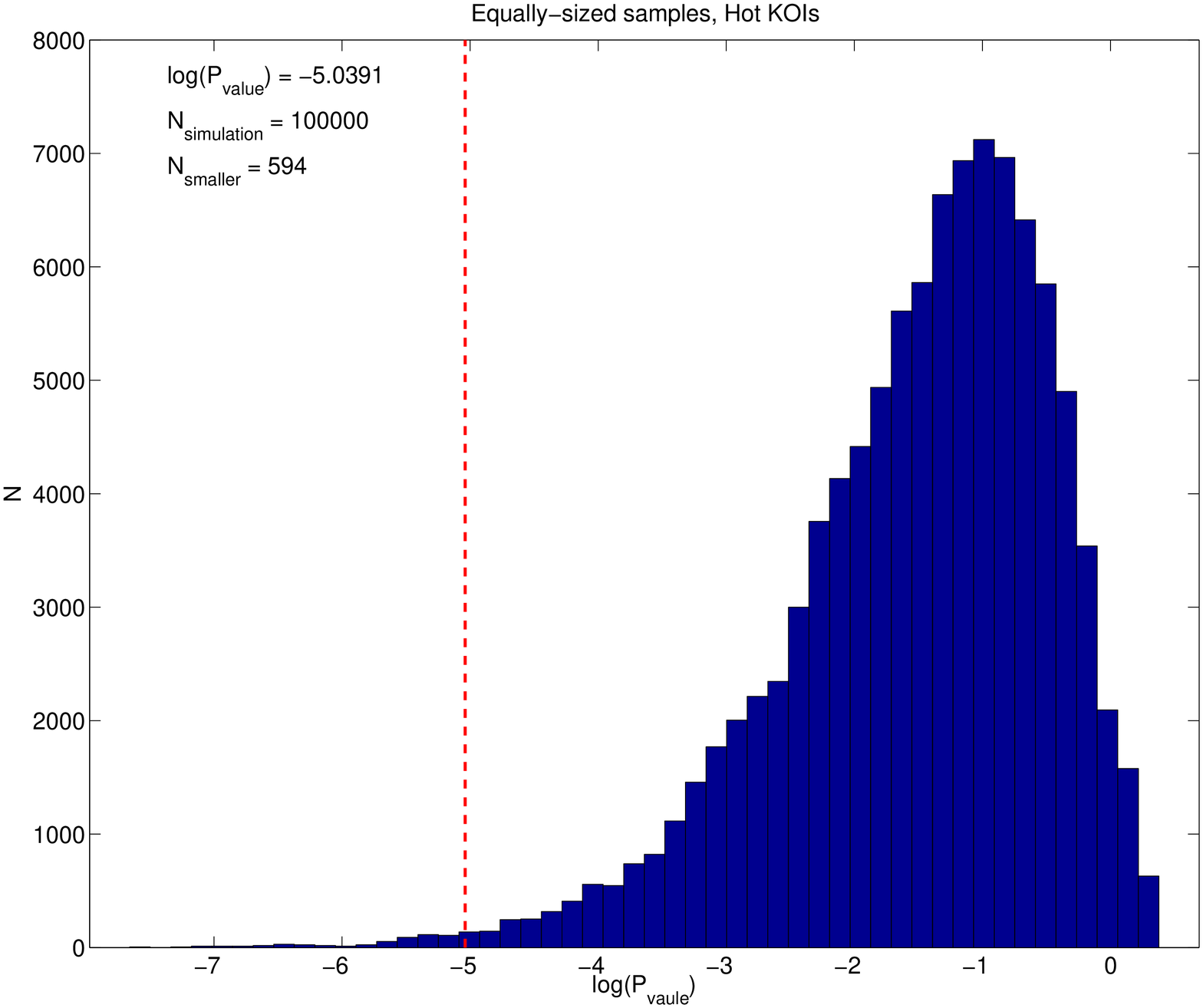}
  \caption{
Histogram of the {\it p}-values of the K-S test, when 100,000 simulated samples of planets around hot stars are compared with the amplitudes of the whole sample of hot stars with detected rotations. Also plotted is the value of the real sample (in dashed red line). 
}
 \label{fig:SimulationKSHot}
\end{figure*}

We performed 100,000 simulations, each of which produced a simulated sample of hot stars with detected rotational periods and amplitudes, with enough S/N so the planets could have been detected. To compare the simulated sample with the real one we used two metrics---the median of the amplitudes and the {\it p}-value of the two-sample K-S test when compared with the whole sample of hot stars. Histograms of the 100,000 derived two measures are plotted in Figures~\ref{fig:SimulationAmpMedianHot} and \ref{fig:SimulationKSHot}.

%----------------------------------------------------------------
%Figure 4
%
%----------------------------------------------------------------

%----------------------------------------------------------------
%Figure 5

%---------------------------------------------------------------

The histogram of Figure~\ref{fig:SimulationAmpMedianHot}, centered around log-amplitude of $\sim3.15$ (in ppm units), as opposed to the median of the hot stars at $\sim3.25$, corroborates the suspicion of the referee---the detection threshold of planets results indeed in smaller amplitudes. The difference in log-amplitude indicates that this effect reduces the amplitudes by a factor of $\sim0.8$. However, the figure shows further that the median of the real sample of hot KOIs is smaller by another factor of 0.8. This might suggest that the smallness of the amplitudes of the hot KOIs could be due {\it also} to some geometrical distribution. 

How significant is this finding? The simulations presented in the two figures show that out of the 100000 simulations, only 594 showed a {\it p}-value of the K-S test smaller than the real one, and only 102 had a smaller median. 
The implication is that  the chance of observing such a difference by a statistical fluke is 
0.6\%, according to the {\it p}-value of the K-S test, or 0.1\%, if we adopt the median 
metric. Figure~\ref{fig:SimulationKSHot} shows that the distribution of the K-S {\it p}-values has a long tail toward small values, if plotted on the log scale, and therefore might not be the best metric to distinguish between real and fluke result. We tend to adopt the 0.1\% result, and contend that the smallness of the hot KOI amplitudes, by a factor of 0.8, is detected with a high statistical significance.  

\clearpage

%======================================
% Section 6
\section{The Geometry of the Spin--Orbit Inclinations of the KOIs}
\label{sec:geometry}
%======================================

The analysis presented here might have some implication on the geometrical distribution of the spin--orbit inclinations of the KOIs.

Suppose that the observed amplitude of the photometric spot modulation of a given star is proportional to $\sin i_{\rm rot}$, where $i_{\rm rot}$ is the inclination of the axis of rotation of that star relative to our line of sight \citep[e.g.,][]{jackson12}. In such a case, we might expect naively that the average amplitude of a {\it sample} of observed stars should be proportional to the expected value of $\sin i_{\rm rot}$ of the sample---$<\sin i_{\rm rot}>$ \citep[][]{jackson13}. 
In real cases we have to consider the convolution of the distribution of the amplitudes with the distribution of inclinations, and take into account the detection threshold of the analysis, two factors that complicate the discussion. However, we assume these factors do not change the general picture we are trying to draw here.   

Now consider two samples, one with random orientations of  its stellar rotational axes in space, with 
$<\sin i_{\rm rot}>=\pi/4$, and the other sample with all stars having $i_{\rm rot}\simeq 90^{\circ}$, all other features being equal.  We then expect the ratio of the two amplitude averages of the two samples to be of the order of $4/\pi\simeq1.25$. 

This is indeed close to what we get for the cool stars, when we compare the rotational amplitudes of the KOIs and those of the single stars of the \kepler\ mission. We do find that the median of the amplitudes of the cool KOIs is larger by a factor of $\simeq1.25$ than the median of the single star amplitudes. If we assume that the spin axes of the single stars are randomly oriented, then our analysis suggests that the stellar spin of the KOIs  have their rotational inclination close to $90^{\circ}$. 
This is consistent with the assumption that the spin axes of the cool  KOIs are 
 aligned with the angular momentum of their transiting planets, which must have their {\it orbital} inclination close to $90^{\circ}$. 

An opposite picture is revealed for the hot KOIs.
The median amplitude of the hot KOIs is probably {\it smaller} by  the same factor, $\sim1.25$, when compared to the median of the amplitudes of the single stars, taking into account the detection bias. This also might be accounted for by a geometrical effect. The maximum geometrical suppression of $<\sin i_{\rm rot}>$ due to spin--orbit misalignment is achieved when all rotational axes are orthogonal to the orbital axes, which have their orbital inclinations close to  $90^{\circ}$. This leads to $<\sin i_{\rm rot}>=2/\pi$, while for random orientation 
$<\sin i_{\rm rot}>=\pi/4$. Therefore, the reduction factor of  
$<\sin i_{\rm rot}>$ is about $1/1.25$. 
Obviously, any other geometrical spin--orbit distribution 
\citep[e.g.,][]{morton14} should result in a less dramatic reduction. Again, the convolution with the
distribution of the amplitudes and the detection bias are going to blur the picture even more. Nevertheless, the fact that the simplest approach yields a factor similar to the one inferred from the analysis is encouraging.

%-----------------------------------------------------------
We wish to call attention to two caveats of the analysis presented here:
\begin{itemize}
\item
Our analysis is not sensitive to the difference between prograde
and retrograde rotational inclinations, as the amplitude depends on the absolute value of $\sin i_{\rm rot}$. Namely, the cool KOIs, for which our analysis suggests rotational axes aligned with the planetary orbital motion, could have prograde or retrograde motion. Different approaches, like the one suggested by \citet{mazeh14}, can shed some light, admittedly for only a very small number of cases, on this question.

\item

We tried to look for differences between large and small planets, but could not convince ourselves, after the referee's doubts in particular, that the results are real, and not due to some observational biases. We suspect that the indirect approach of the present work is not capable of distinguishing between the large and small planets.

\end{itemize}

%====================
%Section 6

\section{Short and Long Orbital Periods around Cool KOIs}
\label{sec:period}
%====================

\begin{figure*}[h!]
 \centering
 \includegraphics[width=1.0\linewidth]{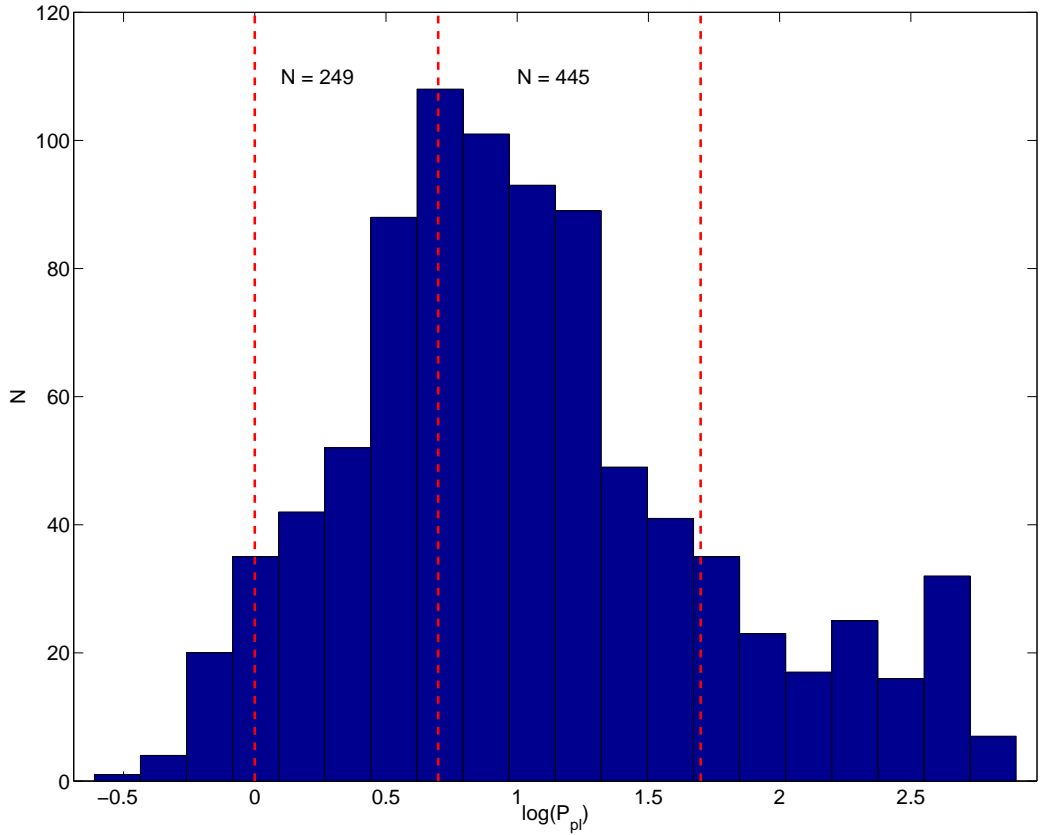}
  \caption{
Histogram of the orbital periods of the innermost planets of the cool KOIs with detected photometric rotation. The three dashed lines show the two period ranges compared here.
}
 \label{fig:PplHistCool}
\end{figure*}

After showing that the larger photometric amplitudes of the cool KOIs might be an indication for low obliquity of their planets, it is of interest to find out if this phenomenon is limited only to KOIs with  short orbital periods, or it extends to the long periods. We therefore divided the cool KOIs into two period ranges, somewhat arbitrarily, of {\it innermost} orbital periods of  1--5 days and 5--50 days, as indicated in the orbital-period histogram, plotted in Figure~\ref{fig:PplHistCool}. Planets with orbital periods outside the range 1--50 days might be too close or too far away from their parent stars to show the general trend we wish to study.

%----------------------------------------------------------------
%Figure 6

%---------------------------------------------------------------

%----------------------------------------------------------------
%Figure 7

\begin{figure*}[h!]
 \centering
 \includegraphics[width=1.0\linewidth]{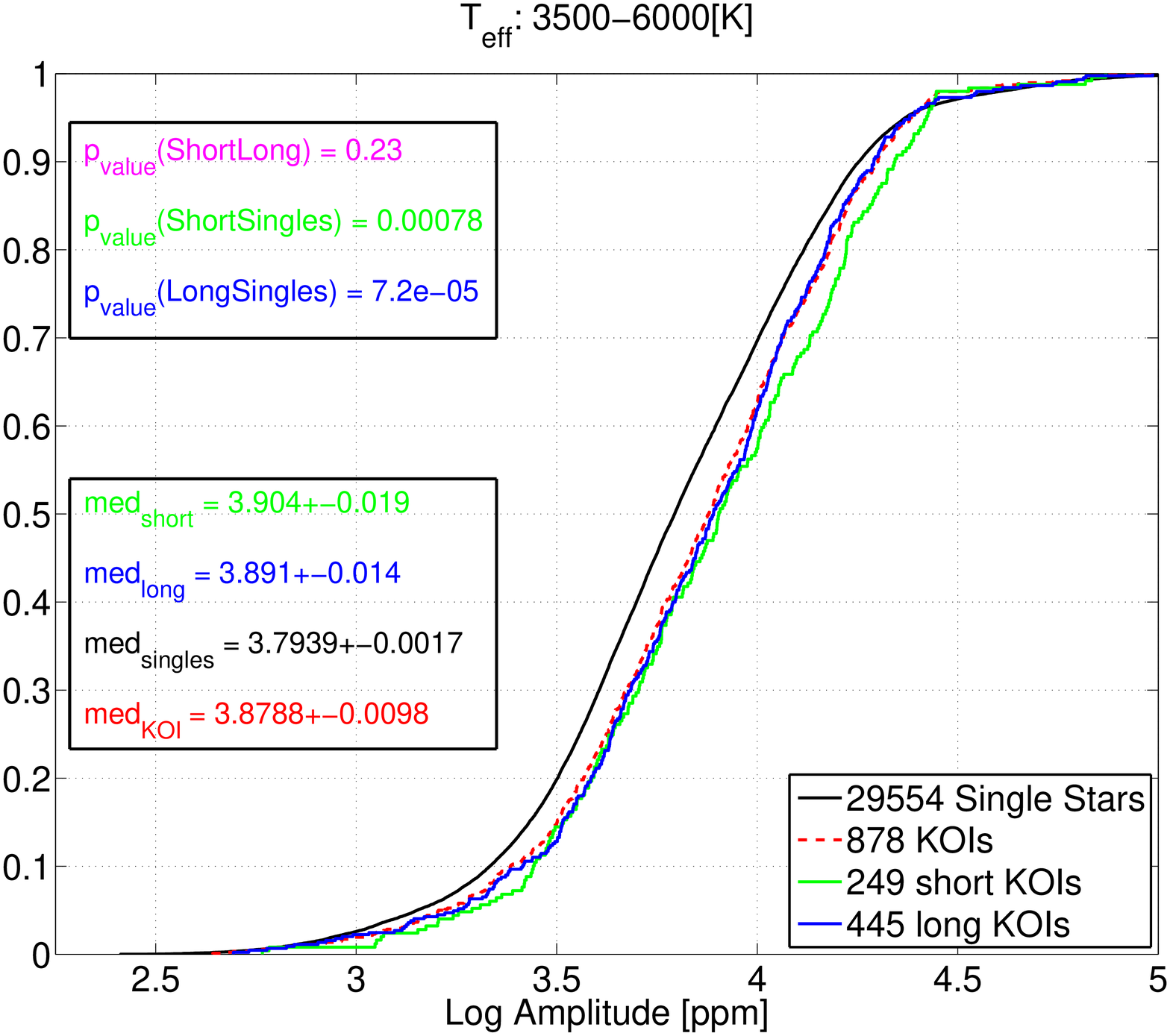}
  \caption{
Cumulative distributions of the photometric amplitudes of the cool KOIs with short (1--5 days) and long (5--50 days) orbital periods. Distributions of all cool KOIs and \kepler\ single stars (Figure~\ref{fig:AmpKS}, left panel) are given for reference.  The medians and the {\it p}-value of the two-sample K-S test for the different samples are given too. 
}
 \label{fig:ShortLongPplKS_Cool}
\end{figure*}
%---------------------------------------------------------------

In Figure~\ref{fig:ShortLongPplKS_Cool} we plot the
cumulative distributions of the photometric amplitudes of the cool KOIs with short (1--5 days) and long (5--50 days) orbital periods. Distributions of all cool KOIs and \kepler\ single stars are given for reference.  The medians and the {\it p}-value of the two-sample K-S test for the different samples are given too. 
The figure shows that the two subsamples have very similar amplitudes, with almost identical medians,  distinctively different from the single cool stars. This suggests that the feature we have discovered---the larger photometric rotational amplitudes of the cool KOIs, is not only restricted to short orbital periods, but extend up to periods of 50 days. 
In our suggested interpretation, the low obliquity of planets around cool stars extends up to at least 50 days.

%====================
%Section 8

\section{Summary and Discussion}
\label{sec:summary}
%====================

In this study we compare the amplitude distribution of the rotation of 993 KOIs with the amplitudes of 33,614 single stars observed by the \kepler\ mission in  temperature range of 3500--6500\,K, in order to shed some light on the statistical distribution of spin--orbit obliquity of the planetary systems.

The analysis presented here is rather limited, because (1) the observed distribution of the amplitudes of any KOI sample is the result of the convolution of the true obliquity with the real amplitude distribution of the sample, (2) we are sensitive only to the line of sight inclination and not to the true obliquity, and (3) our measurements are subject to observational biases. The latter is stronger for smaller planets and hot stars. Although we can try and debias the observations, this inserts a high degree of uncertainty into our results. On the other hand, the analysis presented here does not require any additional observational resources, and can be applied to a large sample of KOIs, with short and long orbital periods alike.

The comparison between the amplitude distributions shows that in the temperature range of 3500--6000\,K the amplitudes of the KOIs are larger, on the order of 10\%, 
than those of the single stars, consistent with the assumption that the spin--orbit obliquity is rather small for cool stars. The low obliquity extends up to at least 50 days.
For the hot stars, with a temperature range of 6000--6500\,K, we have found the opposite effect.
The amplitudes of the hot KOIs are probably {\it smaller} by the same factor, of the order of 10\%, when compared to the amplitudes of the single stars, taking into account the detection bias. This could be the result of the high obliquity of the hot stars. Our findings are consistent with \citet{winn10} and \citet{albrecht12}, who discovered such a transition between aligned and non-aligned systems  at about 6250\,K.  

The analysis presented here might shed some light on the mechanism responsible for the spin--orbit dichotomy between planets around cool and hot stars. In their seminal paper, \citet{winn10} discussed two possible  scenarios to account for their findings. The first had to do with formation mechanism ---
planets around hot stars might have originally formed with
high obliquities, while planets around low-mass stars have been 
formed with low obliquities \citep[see also][]{lai14,batygin12}.
\citet{winn10} then focused on a different scenario, where the
spin--orbit dichotomy arises from a later evolution of the system. In
their scenario planets go through a high obliquity phase,
probably due to planet--planet scattering, which is then followed by
a long tidal interaction period. The tidal interaction affects hot
and cool stars differentially. \citet{winn10} suggested that since cool stars
have convective envelopes, their tidal interaction with the planets
is more effective, leading to the alignment of the 
planetary orbit with the rotation of its host cool star. Conversely, hot stars,
whose envelopes are radiative, cannot align the planetary orbits. 
 
Within the tidal realignment models, the
alignment timescale depends on the planet's separation
from its host star. Such models therefore predict that only short-period planets should be aligned with their
host star's rotation, as the tidal realignment becomes negligible
for long-period planets. 
Our results are probably not aligned with this prediction, because we find evidence that the small obliquity of  the cool KOIs extends to at least 50 days, while the tidal planet--star interaction is probably not very effective in long, 10--50 days, orbital periods. 

\kepler\ provided us with the rare capability of studying a sample of about 150,000 almost uninterrupted LCs, out of which we discovered thousands of new planets and tens of thousands stellar rotations. The combination of the two can be used as a new tool to explore the characteristic statistical features of planet--star interaction, of which this study is an example. To better exploit this exciting opportunity, investigating the planet--star obliquity in particular, we should combine different techniques, like R-M effect, the line-broadening approach, asteroseismology, and spot crossing analysis, applied to specific systems, together with statistical approaches like the one presented here. This could reveal more features of the obliquity and its dependence on stellar temperature and planetary mass and  orbital period, and shed some light on the formation and evolution of planetary systems.  
 
%--------------------------------------------------------

\acknowledgments
\noindent {\it Acknowledgments:}
We are indebted to the referee for his/her very thoughtful advice that substantially improved the previous version of the paper. We wish to deeply thank Tomer Holczer for his tests to
identify EB contamination, Gil Nachmani, Suzanne Aigrain and Simchon Faigler for illuminating and helpful discussions, and Eric Agol for useful comments.
The research leading to these results has received funding to T.M.\ from
the European Research Council under the EU's Seventh Framework
Programme (FP7/(2007-2013)/ ERC Grant Agreement No.\,291352).
T.M.\ also acknowledges support from
the Israel Science Foundation (grant No.~1423/11).
T.M.\ and H.B.P.\ acknowledge support from the Israeli Centers of
Research Excellence (I-CORE, grant No.~1829/12). H.B.P.\ acknowledges
support from the European Union's---Seventh Framework Programme (FP7/2007-2013{]})
under grant agreement no.~333644\textendash{}MC\textendash{}GRAND,
as well as the ``Minerva center for life under extreme planetary
conditions''.
T.M. is grateful to the Jesus Serra  Foundation that enabled an extended visit to the 
Instituto de Astrofísica de Canarias, and to his hosts, Hans Deeg and Rafael Rebolo in particular. The last phase of this work has been completed during this visit. 
This
research has made use of the NASA Exoplanet Archive, which is
operated by the California Institute of Technology, under
contract with the National Aeronautics and Space Administration
under the Exoplanet Exploration Program. All of the data
presented in this paper were obtained from the Mikulski Archive
for Space Telescopes (MAST). STScI is
operated by the Association of Universities for Research in
Astronomy,
Inc., under NASA contract NAS5-26555. Support for MAST for
non-{\it HST}
data is provided by the NASA Office of Space Science via grant
NNX09AF08G and by other grants and contracts.

%\bibliography{koi_bib}
%\bibliographystyle{apj}

%\begin{comment}

\end{document}